\documentclass[a4paper,12pt]{article}
\usepackage{amsmath}
\usepackage{amssymb}

\usepackage{authblk}
\usepackage{color}

\newcommand{\calpha}{ \alpha_r({\bf x},{\bf p}) }
\newcommand{\calphax}{ \alpha_1(0,{\bf p}) }
\newcommand{\calphay}{ \alpha_2(0,{\bf p}) }
\newcommand{\pol}{{\mathbfcal E}_r({\bf p}) }
\newcommand{\modp}{|{\bf p}|}

\newcommand{\BbbC}{\mathbb{C}}

\DeclareMathOperator{\sgn}{sgn}

\DeclareMathOperator{\Realpart}{Re}
\DeclareMathOperator{\Imagpart}{Im}

\DeclareMathOperator{\si}{si}
\DeclareMathOperator{\Si}{Si}
\DeclareMathOperator{\Ci}{Ci}
\DeclareMathOperator{\Cin}{Cin}

\numberwithin{equation}{section}

\DeclareMathAlphabet\mathbfcal{OMS}{cmsy}{b}{n}

\title{Can a localised quantum system\\ see soft photons?}

\author{Sanved Kolekar$^{1}$\footnote{sanved.kolekar@cbs.ac.in} \,}
\author{Jorma Louko$^{2}$\footnote{jorma.louko@nottingham.ac.uk}}

\affil{$^{1}$ UM-DAE Centre for Excellence in Basic Sciences, \\  Mumbai 400098, India}

\affil{$^{2}$ School of Mathematical Sciences, 
University of Nottingham,\\ 
Nottingham NG7 2RD, 
UK}

\date{August 2021}
\begin{document}
\maketitle


\begin{abstract}
We ask whether soft photons, defined by asymptotic charges, can have consequences for the outcome of localised quantum processes. We consider a spatially localised two-state system, at rest in flat spacetime, 
coupled to a $U(1)$ gauge invariant charged scalar field. 
We find that the system's de-excitation rate does depend on the soft charges that correspond to the radial component of the electric field dressing at the asymptotic infinity; the excitation rate, by contrast, remains zero, 
regardless of the soft charges. Some implications are discussed.
\end{abstract}

\section{Introduction}

Asymptotically flat spacetimes are known to possess an infinite number of asymptotic symmetries at their null infinity, quantified by the Bondi-Metzner-Sachs (BMS) supertranslations~\cite{BMS}. The BMS supertranslation symmetries were shown to be related to both the gravitational memory effect \cite{Zeld} and Weinberg's soft graviton theorem \cite{Weinberg}, to form the \textit{Universal Triad} relation \cite{Strom-bms1,He1,Winicour-memory,Ashtekar-memory,Strom-zhiboedov}; the significance of this was recently realised by Hawking, Perry and Strominger (HPS), who conjectured that applying these relations to an asymptotically flat black hole spacetime implies the existence of an infinite number of soft hairs for the black hole \cite{HPS1}. The BMS supertranslations at the asymptotically flat infinity are accompanied by local super-rotation symmetries, which extend the local Lorentz group~\cite{Barnich}. These super-rotations are connected to a new type of gravitational memory known as the spin-memory effect~\cite{Pasterski1}, and they are related to the subleading term in Weinberg's soft graviton theorem \cite{Cachazo,Kapec1}. 

HPS showed \cite{HPS2} that a Schwarzschild black hole can be implanted with soft hair by an infalling supertranslated null shockwave without spherical symmetry. A similar physical process holds for a Rindler horizon with an infalling supertranslated null shockwave without planar symmetry~\cite{Kolekar1}. A perturbative analysis of the quantum entanglement across the Rindler horizons, in terms of the entanglement monotone negativity, showed that the supertranslational hair 
implanted by the shockwave modulates the entanglement between the opposing Rindler wedges in quantum field theory~\cite{Kolekar2}. For the Schwarzschild black hole case, these results suggest that, within a perturbative treatment, the negativity between an infalling and outgoing Hawking pair should be degraded due to an infalling soft-hair-implanting shockwave, while there should be linear order generation of negativity between two outgoing Hawking particles. 

A corresponding electromagnetic memory effect, of both the ordinary and non-linear type (also referred as null memory), was demonstrated in~\cite{Garfinkle}. Here, charges in a suitable detector at asymptotic infinity receive a kick, that is, the charges retain a residual velocity, instead of a deformation of the detector as in the gravitational memory case. The ordinary memory corresponds to the difference in the radial electric field at future null infinity, while the non-linear type is due to the flux of the massless charges that reach future null infinity, in analogy with the Christodoulou \cite{Chris} non-linear gravitational memory, wherein the gravitational wave memory is due to the flux of gravitational waves at future null infinity. A  corresponding universal \textit{triad} similar to the gravitational triad described above holds \cite{He2, Kapec2, Mohd, Pate, Pasterski2} for the electromagnetic case as well, namely between the large $U(1)$ gauge symmetries at null infinity, the electromagnetic memory effect and Weinberg's soft photon theorem~\cite{Weinberg}. The large $U(1)$ gauge  symmetries spontaneously break the degenerate vacua with the soft photons as the corresponding Goldstone mode. The Ward identities associated with the large $U(1)$ gauge  symmetries are shown to be related to the leading Weinberg soft factor. 

The analysis in \cite{Garfinkle} of the electromagnetic memory effect considered only the $E$ type radiation which leads to the residual drift. A new type of electromagnetic memory, having a form similar to the Aharonov-Bohm effect, was demonstrated in~\cite{Mao}, wherein a position displacement is induced for a charged particle due to B type mode radiation. In analogy with the gravitational case, the new type of electromagnetic memory is equivalent to the subleading terms in the soft photon theorem. An experimental setup using superconducting nodes connected to Josephson junctions on a sphere was proposed by Susskind to measure the electromagnetic memory effect~\cite{Susskind}. In~\cite{Bachlechner}, an experimental proposal to test the electric Aharonov-Bohm effect using Josephson junction to measure the relative phase shift in the due to the potential difference between two superconductors is discussed.

In this paper we ask whether one could observe, in principle, effects due to soft photons in suitably \emph{localised\/} quantum processes. Based on the universal triad relations mentioned above, such an outcome can then be interpreted as a form of the electromagnetic memory effect. 
Concretely, we consider a two-level quantum detector whose charged monopole moment couples linearly to a $U(1)$ gauge-invariant complex scalar field in a flat spacetime. The gauge invariance of the complex scalar field incorporates a soft photon dressing factor and hence couples the detector also to the electromagnetic field. 

The quantum system under consideration is described in section~\ref{sec:gauge-inv-coupling}. The expressions for the effective Wightman function and the transition rate of the local quantum detector, moving inertially in flat spacetime, in the vacuum state of both the complex scalar field and the electromagnetic field, are found to depend on the soft photon dressing. Section \ref{special} addresses an angle-dependent Coulombic type electric field, and Section \ref{general} addresses the general case of an electric field dressing, described only by its asymptotic properties. In both cases, we find that the detector's excitation rate vanishes, bearing no effect of the soft charges, but the de-excitation rate depends on the soft charges that correspond to the radial part of the electric field dressing. Section \ref{sec:discussion} gives the conclusions and a brief discussion. Some technical material is deferred to two appendices. 

In asymptotic expansions, ${\cal O}(x)$ denotes a quantity such that ${\cal O}(x)/x$ is bounded as $x\to0$, $o(x)$ denotes a quantity such that $o(x)/x \to 0$ as $x\to0$, and $o(1)$ denotes a quantity that goes to zero in the limit under consideration.

\section{A quantum detector with gauge invariant coupling\label{sec:gauge-inv-coupling}}

In this section we describe a point-like two-level quantum system moving on a time-like trajectory in flat spacetime, coupled linearly to a complex $U(1)$-gauged scalar field. The central input is that we take the detector be coupled to the \emph{gauge-invariantly dressed\/} version of the scalar field, which implies that the detector is indirectly coupled also to the electromagnetic field. We describe the setup below and obtain, within linear perturbation theory, the general expression for the detector's transition rate in terms of the corresponding Wightman function. 

\subsection{Classical fields}

The Lagrangian of the electromagnetic field coupled to a charged scalar field is given by
\begin{equation}
{\cal L} = \dfrac{-1}{4} F_{ab} F^{ab} - |D_a \phi^2| - m |\phi|^2 , 
\end{equation}
where the electromagnetic tensor $F_{ab}$ is defined in terms is the four-vector potential $A_b$ as $F_{ab} = \partial _a A_b - \partial_b A_a $ and the covariant derivative is defined through its action on field $\phi$ through $D_a \phi = \partial_a \phi - iq A_a \phi$.
As is well known, the above action is invariant under the $U(1)$ gauge transformation 
\begin{subequations}
\label{eq:gauge-transf}
\begin{eqnarray}
\phi(x) & \longrightarrow & e^{-iq \Omega(x)} \phi(x) ,  \\
A_a & \longrightarrow & A_a - \partial_a \Omega . 
\end{eqnarray}
\end{subequations} 

The field $\phi$ is not gauge invariant. 
A gauge-invariant field operator can however be constructed by multiplying  
$\phi$ with phase factor that depends on the electromagnetic field, as introduced by Dirac in 1955~\cite{dirac}. The new, `dressed' field operator is given by
\begin{equation}
\Phi(x) = e^{i C(x)} \phi(x) , 
\label{dressedfieldoperator}
\end{equation}
where 
\begin{equation}
C(x) = \int d^4 x^\prime \, f^a(x, x^\prime) A_a(x^\prime) , 
\label{Cdefine}
\end{equation}
and the two-point function $f^a(x, x^\prime)$ that specifies the dressing is a solution to 
\begin{equation}
\partial^\prime_a f^a(x, x^\prime) = q \, \delta^4(x-x^\prime) . 
\label{fconstraint}
\end{equation}
In words, $f^a(x, x^\prime)$ satisfies in its second argument four-dimensional Poisson's equation, sourced by the Dirac Delta distribution. 
Under the gauge transformation \eqref{eq:gauge-transf}, the phase factor $e^{i C(x)}$ transforms as
\begin{eqnarray}
e^{i C(x)} & \rightarrow & e^{i\int d^4 x^\prime \, f^a(x, x^\prime) A_a(x^\prime)} e^{ -i \int d^4 x^\prime \, f^a(x, x^\prime) \partial_a^{\prime} \Omega(x^\prime) }  \nonumber \\
&=& e^{i C(x)} e^{ - i\int d^4 x^\prime \, \partial_a^{\prime} \left[ f^a(x, x^\prime) \Omega(x^\prime) \right]} \; e^{ - i\int d^4 x^\prime \, \left[ \partial_a^{\prime} f^a(x, x^\prime)\right] \Omega(x^\prime) } \nonumber \\
&=& e^{i C(x)} \; e^{ - i q \Omega(x) }, 
\label{invariance}
\end{eqnarray}
where in the second line we have used \eqref{fconstraint} and assumed that $f^a(x, x^\prime) \Omega(x^\prime) \rightarrow 0$ sufficiently fast at infinity. 
The dressed field operator $\Phi(x)$ \eqref{dressedfieldoperator} is hence a gauge invariant charged scalar field operator. 

Note that while the calculation in \eqref{invariance} assumes a sufficient falloff for $f^a(x, x^\prime) \Omega(x^\prime)$ at infinity, the falloff required of $f^a(x, x^\prime)$ depends on what is assumed about~$\Omega$. The dressing construction hence applies not just for `small' gauge transformations, in which $\Omega$ vanishes at infinity, but also for `large' gauge transformations, in which $\Omega$ does not vanish at infinity, provided the falloff of $f^a(x, x^\prime)$ is sufficiently strong~\cite{Kapec2}. This is the case that we shall consider in Sections \ref{special} and~\ref{general}. 

To set the choice of the dressing in its larger context, we recall that in 
a general curved spacetime geometry with an asymptotically flat infinity, 
a large gauge transformation rotates the asymptotic dressed states by a phase that arises from the factor $\int d^4 x^\prime \, \partial_a^{\prime} \left[ f^a(x, x^\prime) \Omega(x^\prime) \right]$. 
The undressed charged particle states without electric field dressing are in fact not the eigenstates of the asymptotic Hamiltonian and hence cannot be used as a basis for asymptotic states~\cite{scattstates}. The additional phase factor plays a crucial role to connect the corresponding Ward identity of the $S$ matrix with the leading soft factors in the soft photon theorem for massive particles \cite{Pate}. In the present, we allow the gauge transformation $\Omega(x^\prime)$ to be arbitrary, that is either small or large, but we consider the electric field configurations $f^{\alpha}(x, x^\prime)$ with suitable falloff conditions for large $r$ on spacelike hypersurfaces such that the vanishing of the combined factor $f^a(x, x^\prime) \Omega(x^\prime) $ at the infinity sets the additional phase factor to zero as shown in~\eqref{invariance}.

\subsection{Quantum fields}

We quantize the system assuming the fields $A_a(x)$ and $\phi(x)$ to be free fields at the lowest order in perturbation theory. 

In standard Minkowski coordinates $(t,{\bf x})$, the charged scalar field $\phi$ has 
the usual mode expansion 
\begin{eqnarray}
\phi(x) &=& \int \dfrac{d^3 \textbf{k}}{(2\pi)^3} \dfrac{1}{\sqrt{2 \omega_{\bf k}}} \left( a_{\bf k} \, e^{i(\textbf{k} \cdot \textbf{x} - \omega_{\bf k} t)} +  b^{\dagger }_{\bf k} \, e^{-i(\textbf{k} \cdot \textbf{x} - \omega_{\bf k} t)}\right), \nonumber \\
\phi^{\dagger}(x) &=& \int \dfrac{d^3 \textbf{k}}{(2\pi)^3} \dfrac{1}{\sqrt{2 \omega_{\bf k}}} \left( b_{\bf k} \, e^{i(\textbf{k} \cdot \textbf{x} - \omega_{\bf k} t)} +  a^{\dagger }_{\bf k} \, e^{-i(\textbf{k} \cdot \textbf{x} - \omega_{\bf k} t)}\right),
\label{phiquantum}
\end{eqnarray}
where $(a^{\dagger}_{\bf k}, a_{\bf k} )$ and $(b^{\dagger}_{\bf k}, b_{\bf k} )$ 
are, respectively, the creation and annihilation operators for the positively and negatively charged particles, 
and $\omega_{\bf k} = |{\bf k}|$.  
The nonvanishing commutators are 
\begin{eqnarray}
\left[  a_{\bf k}, a^{\dagger}_{{\bf k}^\prime} \right] &=& (2\pi)^3\delta^3({\bf k} - {\bf k}^\prime), \nonumber \\
\left[  b_{\bf k}, b^{\dagger}_{{\bf k}^\prime} \right] &=& (2\pi)^3 \delta^3({\bf k} - {\bf k}^\prime). 
\label{phicommutation}
\end{eqnarray}
The vacuum $| 0_{\phi} \rangle$ is the normalised state that satisfies $ a_{\bf k}|  0_{\phi} \rangle = 0$ and $ b_{\bf k}|  0_{\phi} \rangle = 0$.

The electromagnetic potential $A_a$ is quantised in the Coulomb gauge ${\bf \nabla} \cdot {\bf A} = 0$, where the boldface symbol ${\bf A}$ denotes the spatial projection and ${\bf \nabla}$ denotes the spatial nabla. Using the equation of motion for $A_a$, the component $A_0$ can be fixed in terms of ${\bf A}$. The Coulomb gauge then sets the value of $A_0$ to be zero and further restricts only two components of ${\bf A}$ to be independent. The resulting mode expansion can be written as 
\begin{align}
{\bf A}(x) = \int \frac{d^3 \textbf{p} }{(2 \pi)^3} \dfrac{1}{\sqrt{2 \omega_{\bf p}}} \sum_{r = 1}^{2}  {\mathbfcal E}_r({\bf p}) \left( d^{\, r}_{\bf p} \, e^{i(\textbf{p} \cdot \textbf{x} - \omega_{\bf p} t)} +  d^{\, r \,\dagger }_{\bf p} \, e^{-i(\textbf{p} \cdot \textbf{x} - \omega_{\bf p} t)}\right), 
\label{Aquantum}
\end{align}
where $\mathbfcal { E}_r({\bf p})$ with $r\in\{1,2\}$ are two polarization vectors that satisfy 
the transversality and orthonormality conditions 
\begin{subequations}
\label{polarization}
\begin{align}
{\mathbfcal E}_r({\bf p}) \cdot {\bf p} &= 0, 
\\
{\mathbfcal E}_r({\bf p}) \cdot  {\mathbfcal E}_s({\bf p}) &= \delta_{rs}, 
\end{align}
\end{subequations}
and the completeness relation
\begin{equation}
\sum_{r = 1}^2  {\mathcal E}^{\alpha}_r({\bf p}) {\mathcal E}^{\beta}_r({\bf p}) = \delta^{\alpha \beta} - \dfrac{p^\alpha p^\beta}{|{\bf p}|^2}, 
\end{equation} 
where the lowercase Greek letters run over the spatial indices $1,2,3$. 
For convenience of what will follow, we also assume that the polarization vectors satisfy the parity condition \begin{align}
{\mathbfcal E}_r({\bf p}) = {\mathbfcal E}_r(- {\bf p}). 
\end{align}
The photon creation and annihilation operators $(d^{\, r \, \dagger}_k, d^r_k )$ 
satisfy the commutation relations 
\begin{subequations}
\begin{align}
\left[  d^r_{\bf p}, d^{\, s \, \dagger}_{{\bf p}^\prime} \right] 
&= 
(2\pi)^3 \; \delta^{rs} \; \delta^3({\bf p} - {\bf p}^\prime) , 
\\
\left[  d^r_{\bf p}, d^{s}_{{\bf p}^\prime} \right]
&=
\left[  d^{\, r \, \dagger}_{\bf p}, d^{\, s \, \dagger}_{{\bf p}^\prime} \right] = 0. 
\label{Acommutation}
\end{align}
\end{subequations}
It follows that the electric field, given by $E_\alpha = \partial_t A_\alpha$, satisfies 
${\bf \nabla} \cdot {\bf E} = 0$. The vacuum $| 0_A \rangle$ is the normalised state 
that satisfies $d^r_k |  0_A \rangle = 0$.

\subsection{The local quantum detector}

Our localised quantum detector is a spatially pointlike charged two-level system~\cite{Takagi:1986kn}, 
moving in Minkowski spacetime on the trajectory $x^a(\tau)$, parametrised by the proper time~$\tau$. 
The Hilbert space is~$\BbbC^4$, with an orthonormal basis that can be written in tensor product notation as 
$| 0\rangle_- \otimes | 0\rangle_+$, 
$| 1\rangle_- \otimes | 0\rangle_+$, 
$| 0\rangle_- \otimes | 1\rangle_+$
and 
$| 1\rangle_- \otimes | 1\rangle_+$, where the subscript $\pm$ indicates the charge. 

The monopole moment operator $m(\tau)$ 
(the spatially pointlike version of the quantum field operator) 
and its Hermitian conjugate are given by 
\begin{subequations}
\label{detectormoment}
\begin{align}
m(\tau) &=  c_+ e^{- i E \tau} + c_-^\dagger e^{i E \tau} , 
\\
m^{\dagger}(\tau) &=  c_- e^{- i E \tau} + c_+^\dagger e^{i E \tau} , 
\end{align}
\end{subequations}
where the real-valued constant $E$ is the detector's energy gap and the operators 
$c_\pm$ and $c_\pm^\dagger$ act on the charge sector indicated by the subscript, by 
\begin{subequations}
\begin{align}
& c_+^\dagger | 0\rangle_+ = | 1\rangle_+ , 
\ \ 
c_+ | 1\rangle_+ = | 0\rangle_+ , 
\ \  
c_+ | 0\rangle_+ = c_+^\dagger | 1\rangle_+ = 0 , 
\\
& c_-^\dagger | 0\rangle_- = | 1\rangle_- , 
\ \ 
c_- | 1\rangle_- = | 0\rangle_- , 
\ \ 
c_- | 0\rangle_- = c_-^\dagger | 1\rangle_- = 0 . 
\end{align}
\label{eq:det-basis-operators}%
\end{subequations}
We assume $E\ne0$. If $E>0$, $| 0\rangle_\pm$ 
are the ground states and $| 1\rangle_\pm$ are the excited states, 
so that $c_\pm$ are the annihilation operators and $c_\pm^\dagger$ are the creation operators; 
if $E<0$, the roles of the ground and excited states are reversed. 
Allowing either sign of $E$ will allow us to discuss 
both detector excitations and detector de-excitations in a uniform notation. 

We take the detector to couple linearly to the gauge invariant 
scalar field operator $\Phi$~\eqref{dressedfieldoperator}, 
with the interaction picture Hamiltonian 
\begin{equation}
{\mathcal H}_{int} = \lambda \chi(\tau) \left(m(\tau) \Phi(\tau) + m^{\dagger}(\tau) \Phi^{\dagger}(\tau)\right) , 
\label{eq:Hint-def}
\end{equation}
where $\lambda$ is a real-valued coupling constant, $\Phi(\tau)$ is the pull-back of $\Phi(x)$ to the detector's worldline, and the real-valued switching function $\chi$ specifies how the interaction is switched on and off. 
${\mathcal H}_{int}$ is Hermitian, and it preserves the total charge: creation of positive charge in the field $\phi$ is accompanied by the creation of negative charge in the detector, and vice versa. 
Crucially for us, ${\mathcal H}_{int}$ is by construction gauge invariant.

\subsection{Excitation and de-excitation rates}

Before the interaction begins, we prepare the scalar field 
and the electromagnetic field in the usual Minkowski vacuum, denoted by 
\begin{equation}
| 0_M \rangle = | 0_\phi \rangle \otimes |0_A \rangle . 
\label{Minvacuum}
\end{equation}
We prepare the detector in the state $|0_d \rangle := | 0\rangle_- \otimes | 0\rangle_+$. 
This is the ground state if $E>0$ and the fully excited state if $E<0$. 

We wish to find the probability of the detector to have
made a transition when the interaction has ceased. 
Working to linear order in perturbation theory, we consider the detector final state 
\begin{align}
| E_d\rangle  : = \frac{\alpha }{\sqrt{|\alpha|^2 + |\beta|^2}} |1 \rangle_- \otimes  |0\rangle_+  
+ \frac{\beta }{\sqrt{|\alpha|^2 + |\beta|^2}} |0\rangle_- \otimes  |1\rangle_+ , 
\label{eq:det-final-state}
\end{align} 
where $\alpha$ and $\beta$ are complex-valued constants, not both equal to zero. 
The term $| 1\rangle_- \otimes | 1\rangle_+$ 
has been omitted without loss of generality because it cannot be produced by a linear order perturbation. 

Now, the linear order amplitude to find the total system in the state 
$|\phi, A \rangle \otimes | E_d \rangle $ after the interaction has ceased is 
\begin{align}
{\mathcal A}_{amp} & = i \lambda  \langle E_d| \otimes \langle A, \phi | \int d\tau \, {\mathcal H}_{int}(\tau) | 0_M \rangle  \otimes |0_d\rangle \nonumber \\
& =  i \lambda \,  \int d\tau \, \chi(\tau)\; e^{i E \tau} \biggl(\frac{\alpha^* }{\sqrt{|\alpha|^2 + |\beta|^2}} \, \langle A, \phi | \,  \Phi(\tau) \, | 0_M \rangle  \nonumber \\
& \hspace{22ex} + \frac{\beta^* }{\sqrt{|\alpha|^2 + |\beta|^2}} \, \langle A, \phi | \, \Phi^{\dagger}(\tau) \,  | 0_M \rangle \biggr) , 
\end{align}
using \eqref{eq:Hint-def} with \eqref{detectormoment} and~\eqref{eq:det-basis-operators}. 
The probability $P_\chi(E)$ to find the detector in the final state~$| E_d \rangle $, 
regardless the final states of the fields, 
is obtained by summing $|{\mathcal A}_{amp}|^2$ over the final states of the fields, with the outcome 
\begin{align}
P_\chi(E) = \lambda^2 F_\chi(E) , 
\end{align}
where the response function $F_\chi(E)$ is given by 
\begin{align}
F_\chi(E) &= \int d\tau'' \, d\tau' \, \chi(\tau'')\chi(\tau') \, 
e^{-i E \left( \tau'' - \tau' \right)}  
\nonumber \\
& \hspace{4ex}\times  
\biggl( \;  \frac{|\alpha|^2 }{|\alpha|^2 + |\beta|^2} \; \langle 0_M| \, \Phi^{\dagger}(\tau^{\prime \prime}) \,  \Phi(\tau^{\prime}) \, | 0_M \rangle  \nonumber \\
& \hspace{8ex} +  \; \frac{|\beta|^2 }{|\alpha|^2 + |\beta|^2} \, \langle 0_M | \;\Phi(\tau^{\prime \prime}) \Phi^{\dagger}(\tau^{\prime}) \,  | 0_M \rangle \biggr) . 
\label{Fintermediate}
\end{align}
Note that \eqref{Fintermediate} does not contain the matrix elements 
$ \langle 0_M| \, \Phi(\tau^{\prime \prime}) \,  \Phi(\tau^{\prime}) \, | 0_M \rangle $ 
and 
$ \langle 0_M| \, \Phi^{\dagger}(\tau^{\prime \prime}) \,  \Phi^{\dagger}(\tau^{\prime}) \, | 0_M \rangle $
because these matrix elements are vanishing. 

The matrix elements in \eqref{Fintermediate} take the form 
\begin{subequations}
\label{eq:innerproducts1and2}
\begin{align}
\langle 0_M| \, \Phi^{\dagger}(\tau^{\prime \prime}) \,  \Phi(\tau^{\prime}) \, | 0_M \rangle  &= \int \dfrac{d^3 \textbf{k}}{(2\pi)^3} \dfrac{1}{2 \omega_{\bf k}} \, e^{ -i \omega_{\bf k} (t^{\prime \prime} - t^\prime)}  \,e^{i\textbf{k} \cdot (\textbf{x}^{\prime \prime} - \textbf{x}^\prime)} \nonumber \\
& \hspace{6ex} 
\times \langle 0_A | \, e^{-i C(x^{\prime \prime})} \, e^{i C(x^\prime)} \, | 0_A \rangle , 
\label{innerproduct1}
\\[1ex]
\langle 0_M| \, \Phi(\tau^{\prime \prime}) \,  \Phi^{\dagger}(\tau^{\prime}) \, | 0_M \rangle  &= \int \dfrac{d^3 \textbf{k}}{(2\pi)^3} \dfrac{1}{2 \omega_{\bf k}} \, e^{-i  \omega_{\bf k} (t^{\prime \prime} - t^\prime)}  \,e^{i\textbf{k} \cdot (\textbf{x}^{\prime \prime} - \textbf{x}^\prime)} \nonumber \\
& \hspace{6ex} 
\times \langle 0_A | \, e^{i C(x^{\prime \prime})} \, e^{-i C(x^\prime)} \, | 0_A \rangle , 
\label{innerproduct2}
\end{align}
\end{subequations}
using the mode expansions~\eqref{phiquantum}, 
the commutation relations \eqref{phicommutation}
and the definition of the Minkowski vacuum~\eqref{Minvacuum}. 
The primed Minkowski coordinates are evaluated at $\tau'$ 
and the double-primed Minkowski coordinates at~$\tau''$. 
We hence have 
\begin{align}
F_\chi(E) = \int d\tau'' \, d\tau' \, \chi(\tau'')\chi(\tau') \, 
e^{-i E \left( \tau'' - \tau' \right)} \; W_\phi(\tau^{\prime \prime}, \tau^{\prime}) 
\; W_A(\tau^{\prime \prime}, \tau^{\prime}), 
\label{eq:Fchi-expr}
\end{align}
where $W_\phi(\tau^{\prime \prime}, \tau^{\prime})$ 
is the pull-back on the detector's worldline of the usual charged scalar field Wightman function, 
\begin{align}
W_{\phi}(x^{\prime \prime}, x^{\prime}) &=   \langle 0_\phi| \, \phi^{\dagger}(x^{\prime \prime}) \,  \phi(x^{\prime}) \, | 0_\phi \rangle =  \langle 0_\phi| \, \phi(x^{\prime \prime}) \,  \phi^{\dagger}(x^{\prime}) \, | 0_\phi \rangle \nonumber \\
&= \int \dfrac{d^3 \textbf{k}}{(2\pi)^3} \dfrac{1}{2 \omega_{\bf k}} \, e^{-i  \omega_{\bf k} (t^{\prime \prime} - t^\prime)}  \,e^{i\textbf{k} \cdot (\textbf{x}^{\prime \prime} - \textbf{x}^\prime)} , 
\label{Wphi}
\end{align}
and 
$W_A(\tau^{\prime \prime}, \tau^{\prime})$ is the pull-back of the electromagnetic dressing two-point function,  
\begin{align}
W_{A}(x^{\prime \prime}, x^{\prime}) & =
\frac{1 }{|\alpha|^2 + |\beta|^2} 
\biggl(  |\alpha|^2 \;  
\langle 0_A | \, e^{-i C(x^{\prime \prime})} \, e^{i C(x^\prime)} \, | 0_A \rangle 
\nonumber \\
&\hspace{13ex} 
+ |\beta|^2 \; \langle 0_A | \, e^{i C(x^{\prime \prime})} \, e^{-i C(x^\prime)} \, | 0_A \rangle  \biggr). 
\label{Wa}
\end{align}

We now specialise to a detector trajectory that is inertial. 
Without loss of generality, we may take the trajectory to be static in the Minkowski coordinates, 
so that 
\begin{align}
(t, {\bf x}) = (\tau, {\bf 0}) . 
\label{inertialtraj}
\end{align}
We further assume that the two-point function $f^a(x,x^\prime)$ that 
defines the photon dressing phase $C(x)$ \eqref{Cdefine} has the form 
\begin{equation}
f^a(x,x^\prime) = \left( 0, \delta(t -t^\prime) {\tilde f}^\alpha({\bf x}, {\bf x}^\prime) \right), 
\label{fchoice}
\end{equation} 
where 
\begin{equation}
\partial^\prime_\alpha {\tilde  f}^\alpha(x, x^\prime) = q \, \delta^3({\bf x} - {\bf x}^\prime). 
\label{ftconstraint}
\end{equation}
Note that this implies that the constraint \eqref{fconstraint} is satisfied. 
The photon dressing phase 
$C(x)$ \eqref{Cdefine} then becomes 
\begin{equation}
C(x) = \int d^3 x^\prime \, {\tilde f}^\alpha({\bf x}, {\bf x}^\prime) A_\alpha(t, {\bf x}^\prime) . 
\label{Cdiracdefine}
\end{equation}

This choice of the photon dressing phase has two consequences. 
First, a technical consequence is that the 
dressing factor $W_{A}$ \eqref{Wa} is invariant under Minkowski time translations. 
As the scalar Wightman function 
$W_{\phi}$ \eqref{Wphi} is also invariant under Minkowski time translations,
the only time-dependence in the response function $F_\chi(E)$ \eqref{eq:Fchi-expr} 
comes from the switching. 
On passing to the long interaction limit in a controlled way~\cite{birrell,Fewster:2016ewy}, 
the transition probability per unit time, or the transition rate, becomes 
\begin{align}
\frac{dP(E)}{d\tau} = \lambda^2 F(E) , 
\label{transrate}
\end{align}
where the time-independent response function $F(E)$ is 
\begin{align}
F(E) 
= \int_{-\infty}^{\infty} ds \; e^{-i E s} \; W_\phi (s) \; W_A (s) , 
\label{F-infinite}
\end{align} 
and we have written 
\begin{subequations}
\begin{align}
W_\phi(s) &:= W_\phi(s,0), 
\\
W_A(s) &:= W_A(s,0), 
\end{align} 
\end{subequations}
using the time translation invariance. In the rest of the paper we shall 
be working with the transition rate as given by \eqref{transrate} and~\eqref{F-infinite}. 
We recall that the transition is an excitation for $E>0$ and a de-excitation for $E<0$. 

Second, a conceptual consequence of the photon dressing phase \eqref{Cdiracdefine} concerns the physical interpretation. 
Acting on $|0_M \rangle$ with the gauge-invariant field operator $\Phi$ 
creates both a charged particle and an electric field, 
as observed by Dirac~\cite{dirac}: denoting by ${\bf E}$ the electric field operator, 
canonically conjugate to ${\bf A}$, the canonical commutation relations imply 
\begin{eqnarray}
{\bf E} \; \Phi(x)| 0_M \rangle = \Phi(x) \left({\bf E} + {\tilde {\bf f}}({\bf x}, {\bf x}^\prime)  \right) | 0_M \rangle. 
\label{eq:f-interpretation}
\end{eqnarray}
In words, this equation shows that the value of the electric field in the state $\Phi| 0_M \rangle$ differs from the corresponding value in the state $| 0_M \rangle$ by ${\tilde {\bf f}}({\bf x}, {\bf x}^\prime)$. 
Also, note that ${\tilde {\bf f}}({\bf x}, {\bf x}^\prime)$ satisfies \eqref{ftconstraint}, which has the form of the Maxwell equation ${\bf \nabla} \cdot {\bf E} = \rho/\epsilon_0$. We may hence identify ${\tilde {\bf f}}({\bf x}, {\bf x}^\prime) $ as an electric field at point~${\bf x^\prime}$, sourced by a point charge $q$ at source point ${\bf x}(\tau)$. 

In summary, the gauge invariant operator $\Phi$ creates a charged particle together with its accompanying electric field, and is hence a natural operator to appear in the gauge-invariant interaction Hamiltonian~\eqref{eq:Hint-def}: in any physical theory, a charge is always accompanied by its electric field.

\subsection{Photon dressing two-point function}

We now proceed to give a transparent expression to the 
photon dressing two-point function $W_A(x,x^\prime)$~\eqref{Wa}. 

To begin, we define the Fourier transform of ${\tilde {\bf f}}({\bf x}, {\bf x}^\prime)$ in its second argument by 
\begin{equation}
{\tilde {\bf F}({\bf x},{\bf p})} = \int d^3 {\bf x}^\prime \,  e^{i\textbf{p} \cdot \textbf{x}^\prime} \; {\tilde {\bf f}}({\bf x}, {\bf x}^\prime) = {\tilde {\bf F}^*({\bf x},-{\bf p})} , 
\label{momentumspace}
\end{equation}
where the last equality follows because ${\tilde {\bf f}}({\bf x}, {\bf x}^\prime)$ 
is by assumption real-valued. With this notation, we decompose $i C(x)$ as 
\begin{align}
i C(x) = C_+(x) + C_-(x) , 
\label{Ccoherent}
\end{align}
where 
\begin{subequations}
\label{C+-}
\begin{align}
C_+(x) &=   \int \frac{d^3 \textbf{p} }{(2 \pi)^3} \sum_{r = 1}^{2}   \alpha_r({\bf x},{\bf p}) \, d^{\, r \,\dagger }_{\bf p} \,  e^{i \omega_{\bf p} t} , 
\\
C_-(x) &= - \int \frac{d^3 \textbf{p} }{(2 \pi)^3} \sum_{r = 1}^{2}  \alpha_{r} ^*({\bf x},{\bf p}) \, d^{\, r}_{\bf p}\, e^{-i \omega_{\bf p} t} , 
\end{align}
\end{subequations}
and 
\begin{equation}
\alpha_r({\bf x},{\bf p})  = i \dfrac{1}{\sqrt{2 \omega_{\bf p}}} \;  {\mathbfcal E}_r({\bf p}) \cdot {\tilde {\bf F}({\bf x},- {\bf p})} . 
\label{alphacoherent}
\end{equation}
In words, $C_+(x)$ contains only electromagnetic 
field creation operators and $C_-(x)$ contains only electromagnetic field annihilation operators. 
We note in passing that this decomposition allows us to interpret $e^{i C(x)}| 0_A \rangle$ as a coherent state~\cite{klauder-skagerstam-book}. We also note the commutator
\begin{align}
\left[ C_+(x^{\prime}), C_-(x^{\prime\prime}) \right] = \int \frac{d^3 \textbf{p} }{(2 \pi)^3} \sum_{r = 1}^{2}   \alpha_r({\bf x}^{\prime},{\bf p}) \, \alpha^*_r({\bf x}^{\prime\prime},{\bf p}) e^{-i \omega_{{\bf p}} \left(t^{\prime\prime} - t^{\prime} \right) } . 
\label{eq:Cpm-commutator}
\end{align}

We wish to evaluate $W_A(x,x^\prime)$~\eqref{Wa}. 
For the matrix element $\langle 0_A | \, e^{-i C(x^{\prime \prime})} \, e^{i C(x^\prime)} \, | 0_A \rangle$, 
we find 
\begin{align}
\langle 0_A | \, e^{-i C(x^{\prime \prime})} \, e^{i C(x^\prime)} \, | 0_A \rangle 
&= e^{-\frac{1}{2} \left[ C_+(x^{\prime \prime}), C_-(x^{\prime \prime}) \right] 
-\frac{1}{2}\left[ C_+(x^{\prime}), C_-(x^{\prime }) \right] 
+ \left[ C_+(x^{\prime}), C_-(x^{\prime\prime}) \right]}
\notag
\\
&= 
\exp \biggl [  - \frac{1}{2}\int \frac{d^3 \textbf{p} }{(2 \pi)^3} \sum_{r = 1}^{2} \biggl \{  |\alpha_r({\bf x}^{\prime \prime},{\bf p}) |^2 +  |\alpha_r({\bf x}^{\prime},{\bf p}) |^2 \nonumber \\
&  \hspace{9ex} - 2 \alpha_r({\bf x}^{\prime},{\bf p}) \, \alpha^*_r({\bf x}^{\prime\prime},{\bf p}) e^{-i \omega_{{\bf p}} \left( t^{\prime \prime} - t^{\prime} \right) } \biggr  \}   \biggr ] ,  
\label{innerprod11}
\end{align}
using first the Baker-Campbell-Hausdorff formula and then the commutator~\eqref{eq:Cpm-commutator}. 
Proceeding similarly, 
we find that the outcome in \eqref{innerprod11} holds also for 
$\langle 0_A | \, e^{i C(x^{\prime \prime})} \, e^{-i C(x^\prime)} \, | 0_A \rangle$. Collecting, we have 
\begin{align}
W_{A}(x^{\prime \prime}, x^{\prime})  &= \exp \biggl [  - \frac{1}{2}\int \frac{d^3 \textbf{p} }{(2 \pi)^3} \sum_{r = 1}^{2} \biggl \{  |\alpha_r({\bf x}^{\prime \prime},{\bf p}) |^2 +  |\alpha_r({\bf x}^{\prime},{\bf p}) |^2 \nonumber \\
&  \hspace{9ex} - 2 \alpha_r({\bf x}^{\prime},{\bf p}) \, \alpha^*_r({\bf x}^{\prime\prime},{\bf p}) e^{-i \omega_{{\bf p}} \left( t^{\prime \prime} - t^{\prime} \right) } \biggr  \}   \biggr ] . 
\label{Wasimp}
\end{align}
Note that the weights $\alpha$ and~$\beta$, introduced in the definition of the detector final state $| E_d\rangle$~\eqref{eq:det-final-state}, no longer appear in~\eqref{Wasimp}.

\section{Soft electric field dressing with a $1/r^2$ radial profile\label{special}}

In this section we evaluate the detector's transition rate 
with a soft electric field dressing that is purely radial, 
assuming that the magnitude has a radial dependence proportional to~$1/r^2$, 
as would be characteristic of a point charge, 
but allowing the magnitude to have non-trivial angular dependence. 
It is this angular dependence that provides the soft charges. 

We choose the dressing field ${\tilde {\bf f}}({\bf x}, {\bf x}^\prime) $ to be \cite{Giddings}
\begin{equation}
{\tilde {\bf f}}({\bf x}, {\bf x}^\prime) = \frac{q}{4 \pi \epsilon_0} \, \frac{g(\theta^{ A^{\prime}})}{|{\bf x} -  {\bf x}^\prime|^3} ({\bf x} -  {\bf x}^\prime) , 
\label{gf}
\end{equation}
where the function $g(\theta^{ A'})$ depends only on the angular coordinates 
$\theta^A = (\theta, \phi)$ 
of the relative position vector, and it satisfies the normalisation condition 
$\int_0^\pi \int_0^{2 \pi} g(\theta^{ A}) \sin{\theta} \, d\theta \, d\phi = 4 \pi $. 
When interpreted as an electric field according to \eqref{eq:f-interpretation}, 
the $1/r^2$ falloff of the magnitude guarantees that the total energy 
gets a finite contribution from the neighbourhood of the infinity.

Soft charges for this configuration can be defined either at null infinity ${\cal I}^{\pm}$ or at spatial infinity $i^0$. Given a weight function $\epsilon(\theta^A)$, where $\theta^A$ refer to the transverse angular coordinates in the advanced (or retarded) Bondi coordinates, the corresponding soft charge is 
\begin{align}
Q_{\epsilon}  = & \int d\Omega \; \epsilon(\theta^A) \lim\limits_{r \to \infty} \left( r^2 E^r \right) \nonumber \\
= & \int d\Omega \; \epsilon(\theta^A) \; g(\theta^{ A}) , 
\label{softcharge}
\end{align}
where the integral over $d\Omega$ runs over the transverse angular coordinates and $E^r$ is the radial component of the electric field.

Now, using \eqref{momentumspace}, the momentum space representation becomes, setting ${\bf x} = 0$, 
\begin{eqnarray}
{\tilde {\bf F}({\bf 0},{\bf p})} &=&  \int d^3 {\bf x}^\prime \,  e^{i\textbf{p} \cdot \textbf{x}^\prime} \; \frac{q}{4 \pi \epsilon_0} \, \frac{g(\theta^{ A^{\prime}})}{|{\bf 0} -  {\bf x}^\prime|^3} ({\bf 0} -  {\bf x}^\prime) . 
\label{gfmomentumspace}
\end{eqnarray}
To evaluate the volume integral $d^3 {\bf x}^\prime$, 
we orient the momentum vector ${\bf p}$ direction to lie along the ${\bf z}$ direction and use spherical coordinates to split the integrals into radial and angular integrals. Collecting the radially dependent terms, the radial integral can be written as
\begin{align}
I_R & = \int_0^\infty dr \, r^2 \, e^{i |{\bf p}| r \cos{\theta}} \times \frac{q}{4 \pi \epsilon_0 r^2}  \nonumber \\
& = \lim_{\epsilon \rightarrow 0^+} \int_0^\infty dr  \, e^{i |{\bf p}| r \cos{\theta}}\times e^{-\epsilon r}  \times \frac{q}{4 \pi \epsilon_0} \nonumber \\
&= \left \{ i \, {\cal P} \left( \frac{1}{|{\bf p}| \cos{\theta}}\right) + \pi \,\delta \left( \modp \cos{\theta} \right) \right \} \frac{q}{4 \pi \epsilon_0} , 
\label{radialintegral}
\end{align}
where the limit $\epsilon\to0^+$ in the second line encodes the distributional interpretation of the integral, 
and in the last line 
${\cal P}$ stands for the Cauchy principal value. 
We hence have ${\tilde {\bf F}(0,{\bf p})} = {\tilde {\bf F}_{{\cal P}}(0,{\bf p})} + {\tilde {\bf F}_{\delta}(0,{\bf p})}$, 
where 
\begin{subequations}
\begin{align}
{\tilde {\bf F}_{{\cal P}}(0,{\bf p})} &= 
\frac{ - i q}{4 \pi \epsilon_0 |{\bf p}|} \, {\cal P} \int_0^{\pi} \int_0^{2 \pi}  \sin{\theta} \, d\theta \, d\phi \; \left( \frac{1}{\cos{\theta}}\right) g(\theta,\phi) \hat{\textbf{r}} , 
\label{Fp-def}
\\
{\tilde {\bf F}_{\delta}(0,{\bf p})} & = \frac{  - q\pi}{4\pi \epsilon_0 |{\bf p}|} \, \int_0^{\pi} \int_0^{2 \pi}  \sin{\theta} \, d\theta \, d\phi \; g(\theta,\phi) \; \delta \! \left( \cos{\theta}\right) \hat{\textbf{r}} . 
\label{Fdelta-def}
\end{align}
\end{subequations}

For ${\tilde {\bf F}_{{\cal P}}(0,{\bf p})}$, we decompose \eqref{Fp-def} into its Cartesian components as 
\begin{align}
{\tilde {\bf F}_{{\cal P}}(0,{\bf p})} = \frac{ - i q}{4 \pi \epsilon_0 |{\bf p}|}\left[\,  Q_{\epsilon_1} \hat{x} + Q_{\epsilon_2} \hat{y} + T_z \hat{z} \, \right] . 
\label{Fp}
\end{align}
Recall that we have chosen ${\bf p}$ to be in the $\hat{z}$ direction, and the polarisation vectors $\pol$ are hence in the $xy$-plane. It follows that $T_z$ does not contribute to $\calpha$~\eqref{alphacoherent}. The quantities that do contribute to $\calpha$ are $Q_{\epsilon_1}$ and $Q_{\epsilon_2}$, which have the expressions 
\begin{subequations}
\label{softcharge1}
\begin{align}
Q_{\epsilon_1} &= {\cal P} \int_0^{\pi} \int_0^{2 \pi}  \sin{\theta} \,  d\theta \, d\phi  \; g(\theta,\phi) \times \left( \cos{\phi}\,  \tan{\theta} \right) , \\
Q_{\epsilon_2} &= {\cal P}  \int_0^{\pi} \int_0^{2 \pi}  \sin{\theta} \,  d\theta \, d\phi  \; g(\theta,\phi) \times \left( \sin{\phi}\,  \tan{\theta} \right) . 
\end{align}
\end{subequations}
Comparing \eqref{softcharge1} and~\eqref{softcharge}, we can identify $Q_{\epsilon_1}$ and $Q_{\epsilon_2}$ to be the soft charges corresponding to the respective transverse functions $\epsilon_1 = {\cal P} \cos{\phi}\,  \tan{\theta}$ and $\epsilon_2 = {\cal P} \sin{\phi}\,  \tan{\theta}$. Note that the Cauchy principal value 
renders the soft charges well defined and finite, 
despite the singularity of $\epsilon_1$ and $\epsilon_2$ at $\theta = \pi/2$. 

For ${\tilde {\bf F}_{\delta}(0,{\bf p})}$, proceeding similarly from \eqref{Fdelta-def} gives 
\begin{align}
{\tilde {\bf F}_{\delta}(0,{\bf p})} 
= \frac{ - q}{4 \pi \epsilon_0 |{\bf p}|}\left[\, Q_{\epsilon_3} \hat{x} + Q_{\epsilon_4} \hat{y} + T^\prime_z \hat{z} \, \right] , 
\label{Fdelta}
\end{align}
where $T_z^\prime$ does not contribute to~$\calpha$, 
while the quantities $Q_{\epsilon_3}$ and $Q_{\epsilon_4}$ that do contribute to $\calpha$ have the expressions 
\begin{subequations}
\label{softcharge2}
\begin{align}
Q_{\epsilon_3} &=  \pi \int_0^{\pi} \int_0^{2 \pi}  \sin{\theta} \,  d\theta \, d\phi  \; g(\theta,\phi) \times \left \{ \cos{\phi}\, \delta \! \left( \theta - \frac{\pi}{2} \right) \right \} 
\notag 
\\ 
&= 
\pi \int_0^{2 \pi}  d\phi  \; g(\pi/2,\phi) \cos{\phi} , 
\\
Q_{\epsilon_4} &=  \pi \int_0^{\pi} \int_0^{2 \pi}  \sin{\theta} \,  d\theta \, d\phi  \; g(\theta,\phi) \times \left \{ \sin{\phi}\, \delta \! \left( \theta - \frac{\pi}{2} \right) \right \} 
\notag 
\\
&= 
\pi \int_0^{2 \pi}  d\phi  \; g(\pi/2,\phi) \sin{\phi} . 
\end{align}
\end{subequations}
Comparing \eqref{softcharge2} with~\eqref{softcharge}, we can identify $Q_{\epsilon_3}$ and $Q_{\epsilon_4}$ as the soft charges corresponding to the respective transverse functions 
$\epsilon_3 =  \pi \cos{\phi}\, \delta \! \left( \theta - \pi /2 \right)$ and $\epsilon_4 = \pi \sin{\phi}\, \delta \! \left( \theta - \pi /2 \right)$. 

Without loss of generality, we may orient the polarisation vectors 
$\pol$ so that 
${{\mathbfcal E}_1({\bf p}) }$ is in the $\hat{x}$ direction and ${{\mathbfcal E}_2({\bf p}) }$ 
is in the $\hat{y}$ direction. 
Substituting \eqref{Fp} and \eqref{Fdelta} in~\eqref{alphacoherent}, we obtain 
\begin{subequations}
\label{eq:calphaxy}
\begin{align}
\calphax & = \frac{q}{4 \pi \epsilon_0 } \; \left( \frac{1}{\sqrt{2\omega_{{\bf p}}}} \right)\;  \frac{1}{|{\bf p}|} \left[\, Q_{\epsilon_1}  - i \, Q_{\epsilon_3}   \, \right] , \\
\calphay & = \frac{q}{4 \pi \epsilon_0 } \; \left( \frac{1}{\sqrt{2\omega_{{\bf p}}}} \right) \;  \frac{1}{|{\bf p}|} \left[\, Q_{\epsilon_2}  - i \, Q_{\epsilon_4}   \, \right] . 
\end{align}
\end{subequations}

Recall again that above we chose ${\bf p}$ to be in the $\hat{z}$ direction, 
and the formulas \eqref{softcharge1} and \eqref{softcharge2} 
for the charges $Q_{\epsilon_i}$ were written with this choice. 
For the response function~\eqref{F-infinite}, we need the pull-back 
of the two-point function $W_{A}(x^{\prime \prime}, x^{\prime})$ \eqref{Wasimp} to the inertial trajectory~\eqref{inertialtraj}, which involves an integration over~${\bf p}$. 
Using \eqref{eq:calphaxy}, we find 
\begin{align}
W_{A}(\tau^{\prime \prime}, \tau^{\prime})  &= \exp \biggl [ - \frac{q^2}{4 \pi \epsilon_0^2 }  \biggl \{ \langle Q^2_{\epsilon_1} \rangle + \langle Q^2_{\epsilon_2} \rangle +  \langle Q^2_{\epsilon_3} \rangle + \langle Q^2_{\epsilon_4} \rangle \biggr  \}  \nonumber \\
&\hspace{10ex}
\times  \int \frac{d \modp }{(2 \pi)^3} \left( \frac{1}{2\omega_{{\bf p}}} \right) \left( 1 -  e^{-i \omega_{{\bf p}} \left(\tau^{\prime \prime} - \tau^{\prime} \right) } \right) \biggr ] , 
\label{Wainertial}
\end{align}
where $\langle Q^2_{\epsilon_i} \rangle$ denotes the average of $Q^2_{\epsilon_i}$ 
over the direction of the unit vector~$\hat {\bf p}$, over the full solid angle. 
Note that because the soft charges appear squared, 
and in combinations that are rotationally invariant about the ${\bf p}$ vector, 
the outcome is unaffected by any choices made for the polarisation vector conventions. 

We emphasise that what appears in \eqref{Wainertial} is the mean of the \emph{square\/} of the soft charge, 
which is nonvanishing whenever the soft charge is nonvanishing, and not the mean of the soft charge itself, which can vanish for some nonvanishing soft charges. We see that any nonvanishing soft charges inevitably affect the response of the local detector. We may think of this effect as arising from the \emph{noise\/} due to the soft photons. 

The integral over $\modp$ in \eqref{Wainertial} is ultraviolet divergent. To regulate the divergence, we introduce an ultraviolet cut-off $\Lambda>0$. Using the results of Appendix~\ref{A}, we find that the regulated two-point function is given by 
\begin{align}
W_{A}(\tau^{\prime \prime}, \tau^{\prime})  &=   \exp \biggl [  - \frac{q^2}{64 \pi^4 \epsilon_0^2 }
\left( \langle Q^2_{\epsilon_1} \rangle + \langle Q^2_{\epsilon_2} \rangle +  \langle Q^2_{\epsilon_3} \rangle + \langle Q^2_{\epsilon_4} \rangle \right) 
\nonumber \\
&\hspace{10ex} \times  \left( \log{\left[ \epsilon + i (\tau^{\prime \prime} - \tau^{\prime}) \right]} + \gamma + \log{\Lambda} \right)    \biggr ] , 
\label{Wainertial2}
\end{align}
where $\gamma$ is the Euler-Mascheroni constant~\cite{DLMF}, the limit $\epsilon \to 0^+$ is understood, 
and the log denotes the branch that is real for positive argument. 

The remaining ingredient of the response function \eqref{F-infinite} is the pull-back of the scalar field two-point function \eqref{Wphi} to the inertial trajectory, given by \cite{birrell}
\begin{align}
W_{\phi}(\tau^{\prime \prime}, \tau^{\prime})  
=  \frac{-1}{4 \pi^2 \left[(\tau^{\prime \prime} - \tau^{\prime}) - i \epsilon \right]^2} , 
\label{Wphiinertial}
\end{align}
where the distributional limit $\epsilon \to 0^+$ is understood. 
From~\eqref{F-infinite}, the response function is given by 
\begin{align}
F(E)
&= \int_{-\infty}^{\infty} ds \; e^{-i E s} \; W_\phi (s) \; W_A (s) 
\nonumber \\
&= \frac{e^{- \tilde{Q} \gamma}}{4 \pi^2 \Lambda^{\tilde{Q}}} \lim_{\epsilon \rightarrow 0^+} \int_{-\infty}^{\infty} ds \;  \frac{e^{-i E s}}{\left( \epsilon + is  \right) ^{2+ \tilde{Q}}} \; , 
\label{transrateinertial}
\end{align} 
where 
\begin{align}
\tilde{Q} : = \frac{q^2}{64 \pi^4 \epsilon_0^2 }  \biggl \{ \langle Q^2_{\epsilon_1} \rangle + \langle Q^2_{\epsilon_2} \rangle +  \langle Q^2_{\epsilon_3} \rangle + \langle Q^2_{\epsilon_4} \rangle \biggr  \} . 
\end{align}
The integral in \eqref{transrateinertial} can be recognised as
the inverse Laplace transform integral of $1/z^{2+ \tilde{Q}}$ 
with respect to negative~$E$. The final expression for the response function is hence 
\begin{align}
F(E) = \frac{(-E) \, \Theta(-E)}{2 \pi} \times \frac{e^{- \tilde{Q} \gamma} \left( -E/\Lambda\right)^{\tilde{Q}}}{\Gamma(2+ \tilde{Q})}  . 
\label{transrateinertialfinal}
\end{align} 

In the limit $\tilde{Q} \to 0$, the response function $F(E)$ \eqref{transrateinertialfinal} 
reduces to the well-known Minkowski vacuum response without a gauge field~\cite{birrell}, 
\begin{eqnarray}
F_{\text{Mink}}(E) = \frac{(-E) \, \Theta(-E)}{2 \pi} ,  
\end{eqnarray} 
vanishing for excitations and proportional to the energy gap for de-excitations. 
In particular, the ultraviolet cut-off $\Lambda$ disappears in this limit. In the presence of the soft charges, $\tilde{Q} > 0$, the excitation rate is still vanishing, but the de-excitation rate gets modified, being now proportional to a higher power of the energy gap, although with an overall constant that depends on the ultraviolet cut-off. 

We conclude that the soft charges do not induce spontaneous excitations in our local detector, but they do affect the detector's de-excitation rate.

\section{Soft electric field dressing with power-law radial asymptotics\label{general}}

In this section we generalise the analysis of the previous section to soft electric dressings described only through their power-law asymptotics at large and small radii. 

\subsection{Soft dressing asymptotics\label{asymp}}

We wish to consider soft dressings ${\tilde {\bf f}}({\bf x}, {\bf x}^\prime)$ whose magnitude is asymptotically proportional to ${|{\bf x} - {\bf x}^\prime|}^{-2}$, as in~\eqref{gf}, at both small and large $|{\bf x} - {\bf x}^\prime|$, but allowing the direction of ${\tilde {\bf f}}({\bf x}, {\bf x}^\prime)$ to be not necessarily radial, and also allowing the angle-dependences at small and large $|{\bf x} - {\bf x}^\prime|$ to differ. 

To accomplish this, we first expand ${\tilde {\bf f}}({\bf x}, {\bf x}^\prime)$ in vector spherical harmonics \cite{hill,barrera,carrascal} as
\begin{align}
{\tilde {\bf f}}({\bf x}, {\bf x}^\prime) = \sum_{l,m} \left \{ f^Y_{lm}(r^\prime) \, \vec{Y}_{lm}(\theta^\prime,\phi^\prime) + f^{\Psi}_{lm}(r^\prime) \, \vec{\Psi}_{lm}(\theta^\prime,\phi^\prime) +  f^{\Phi}_{lm}(r^\prime) \, \vec{\Phi}_{lm}(\theta^\prime,\phi^\prime) \right  \} , 
\label{vectorexpansion}
\end{align}
where $(r',\theta',\phi')$ are the polar coordinates of the vector 
${\bf r}' = {\bf x}^\prime - {\bf x}$, 
and the radial, polar and azimuthal vector spherical harmonics are respectively defined by 
$\vec{Y}_{lm}(\theta^\prime,\phi^\prime) = Y_{lm}(\theta^\prime,\phi^\prime) \hat{{\bf r}'}$, 
$\vec{\Psi}_{lm}(\theta^\prime,\phi^\prime) = r' \vec{\nabla} Y_{lm}(\theta^\prime,\phi^\prime)$ 
and $\vec{\Phi}_{lm}(\theta^\prime,\phi^\prime) = \hat{{\bf r}'} \times \vec{\nabla} Y_{lm}(\theta^\prime,\phi^\prime)$, 
where $Y_{lm}(\theta^\prime,\phi^\prime)$ are the scalar spherical harmonics~\cite{DLMF}. 

We then assume that the expansion coefficients have large and small $r'$ 
expansions that proceed in integer powers of $r'$, starting as 
\begin{align}
f^S_{lm}(r^\prime)  = 
\begin{cases}
{\displaystyle \frac{\left( A^S_{lm} + {}_{(1)}A^S_{lm} r' + {}_{(2)}A^S_{lm} {r'}^2 + \cdots \right)}{{r'}^2 }}
& \text{for} \ r' \rightarrow 0 , 
\\[3ex]
{\displaystyle \frac{1}{{r'}^2} \! \left( B^S_{lm} + \frac{{}_{(1)}B^S_{lm}}{r'} + \frac{{}_{(2)}B^S_{lm}}{{r'}^2} + \cdots \right) } 
& \text{for} \ r' \rightarrow \infty , 
\end{cases} 
\label{fasymptotics}
\end{align}
where the index $S$ runs over the values $Y$, $\Phi$, and~$\Psi$, and the $A$s and $B$s are constants. 
We shall see that assuming only integer powers in \eqref{fasymptotics} 
gives sufficient control of the subleading terms for the purposes of the detector's response.

\subsection{The response function: preliminaries}

To find the response function for the inertial trajectory~\eqref{inertialtraj}, we need to establish auxiliary results about the asymptotics of the coefficients $\alpha_r(0,{\bf p})$~\eqref{alphageneral}. 

Substituting the vector harmonics expansion 
\eqref{vectorexpansion} in \eqref{momentumspace} and~\eqref{alphacoherent}, we obtain
\begin{align}
\alpha_r(0,{\bf p}) = \frac{i }{\sqrt{2 \omega_p}} \sum_{l,m,r,S} 
\int  dr^\prime  d\Omega^\prime \, r^{\prime 2} \, e^{i p r^\prime \cos{\theta^\prime}}  f^S_{lm}(r^\prime) \, \left( {\bf \vec{S}}_{lm}(\theta^\prime,\phi^\prime) \cdot \pol  \right) , 
\label{alphageneral}
\end{align}
where ${\bf \vec{S}}_{lm}$ denotes the vector spherical harmonics 
$\vec{Y}_{lm}$, $\vec{\Psi}_{lm}$ and $\vec{\Phi}_{lm}$. 

As in Section~\ref{special}, we may suppose first, without loss of generality, that the momentum vector ${\bf p}$ is oriented in the $\hat{z}$ direction, and the polarisation vectors 
$\pol$ are oriented so that 
${{\mathbfcal E}_1({\bf p}) }$ is in the $\hat{x}$ direction and ${{\mathbfcal E}_2({\bf p}) }$ 
is in the $\hat{y}$ direction. Since the vector spherical harmonics 
${\bf \vec{S}}_{lm}(\theta^\prime,\phi^\prime)$ 
are polynomial functions of $\cos \theta$, $\sin\phi$ and $\cos\phi$, 
the angular integral in \eqref{alphageneral} can be written as 
\begin{align}
{\bf \vec{Q}}_{lm}^S( p r^\prime) =  \int_{0}^{\pi} d\theta^\prime \sin \theta^\prime \; e^{i p r^\prime \cos \theta^\prime} \; {\bf \vec{h}}_{lm}^S(\cos \theta^\prime) , 
\end{align}
where ${\bf \vec{h}}_{lm}^S(\cos \theta^\prime)$ is the integral of ${\bf \vec{S}}_{lm}(\theta^\prime,\phi^\prime)$ over~$\phi^\prime$. 
Writing $p r' = v$, 
the asymptotic behaviour of $ {\bf \vec{Q}}_{lm}^S(v)$ at small $v$ is obtained by applying a power series, 
while the asymptotic behaviour at large $v$ proceeds in terms of the form $e^{\pm i v} \, {\bf \vec{h}}_{lm}^S(\pm 1)/ v$, $e^{\pm i v} \, {\bf \vec{h}}^{S\prime}_{lm}(\pm 1)/ v^2$, $\dots$,$e^{\pm i v} \, {\bf \vec{h}}^{S(n)}_{lm}(\pm 1)/ v^{n+1}$, $\dots$ where the prime denotes the derivative with respect to the argument and $(n)$ stands for the $n^{th}$ derivative~\cite{wong}. 
Writing $ {\bf \vec{Q}}_{lm}^S(v) =  {\bf \vec{Q}}_{lm+}^S(v) +   {\bf \vec{Q}}_{lm-}^S(v)$, we have 
\begin{align}
{\bf \vec{Q}}_{lm\pm}^S(v) = 
\begin{cases}
{\bf \vec{a}}_{lm \pm}^S + {}_{(1)}{\bf \vec{a}}_{lm \pm}^S v + {}_{(2)}{\bf \vec{a}}_{lm \pm}^S v^2 + \cdots 
& \text{for} \ v \rightarrow 0, 
\\[1ex]
{\displaystyle \frac{e^{\pm i v }}{v} 
\left( 
{\bf \vec{b}}_{lm \pm}^S 
+ \frac{{}_{(1)}{\bf \vec{b}}_{lm \pm}^S}{v} 
+ \frac{{}_{(2)}{\bf \vec{b}}_{lm \pm}^S}{v^2} 
+ \cdots
\right) 
}
& \text{for} \ v \rightarrow \infty , 
\end{cases}
\label{Qasymptotics}
\end{align}
where ${\bf \vec{a}}_{lm \pm}^S$ and ${\bf \vec{b}}_{lm \pm}^S$ are the leading order coefficients in the expansions, and the dots indicate a pattern that continues in integer powers of~$v$. 

The remaining integral over $r$ in \eqref{alphageneral} is of the form
\begin{align}
 \frac{1}{p} \, {\bf \vec{H}}^S_{\pm lm}( p )  =  \int_0^\infty dr \; r^2 \; f^S_{lm}(r) \;  {\bf \vec{Q}}_{lm \pm}^S( p r) , 
\end{align}
where the $f^S_{lm}(r)$ have the asymptotics~\eqref{fasymptotics}. 
To determine the asymptotics of ${\bf \vec{H}}^S_{\pm lm}( p )$ at large and small $p$, we introduce a positive constant $\mu$ and write 
\begin{align}
{\bf \vec{H}}^S_{\pm lm}( p ) = {\bf \vec{H}}^S_{\pm lm<}( p ) + {\bf \vec{H}}^S_{\pm lm>}( p ) , 
\end{align}
where 
\begin{subequations}
\begin{align}
{\bf \vec{H}}^S_{\pm lm<}( p ) =  p \int_0^\mu dr \; r^2 \; f^S_{lm}(r) \;  {\bf \vec{Q}}_{lm \pm}^S( p r) , \label{H<} \\
{\bf \vec{H}}^S_{\pm lm>}( p ) =  p \int_\mu^\infty dr \; r^2 \; f^S_{lm}(r) \;  {\bf \vec{Q}}_{lm \pm}^S( p r) . \label{H>}
\end{align}
\end{subequations}

Consider first ${\bf \vec{H}}^S_{\pm lm<}( p )$~\eqref{H<}. 
The asymptotic expansion at small $p$ is obtained by doing a Maclaurin expansion in $p$ under the integral in~\eqref{H<}. 
To obtain the asymptotic expansion at large~$p$, 
we use \eqref{fasymptotics} and \eqref{Qasymptotics} to rewrite \eqref{H<} as 
\begin{align}
{\bf \vec{H}}^S_{\pm lm<}( p ) = & \, A_{lm}^S  \int_0^{ \mu p} dv  \;  {\bf \vec{Q}}_{lm \pm}^S(v) 
+ \frac{{}_{(1)}A_{lm}^S}{p}
\int_0^{ \mu p} dv  \left[ v  {\bf \vec{Q}}_{lm \pm}^S( v)  -   {\bf \vec{b}}_{lm \pm}^S e^{\pm i v } \right]
\nonumber \\
& \hspace{2ex}
+  {\bf \vec{b}}_{lm \pm}^S  \int_0^\mu dr \; \frac{\left( r^2  f^S_{lm}(r) - A^S_{lm} \right)}{r} \; e^{\pm i p r } 
\nonumber \\
& \hspace{2ex}
+ \frac{{}_{(1)}{\bf \vec{b}}_{lm \pm}^S}{p}
\int_0^\mu dr \; \frac{\left( r^2  f^S_{lm}(r) - A^S_{lm} - {}_{(1)}A^S_{lm} r \right)}{r^2} \; e^{\pm i p r } 
\nonumber \\
& \hspace{2ex}
+ \frac{1}{p}
\int_0^\mu dr \; \frac{\left( r^2  f^S_{lm}(r) - A^S_{lm} - {}_{(1)}A^S_{lm} r \right)}{r^2} 
\notag \\
& \hspace{7ex}
\times \left[ (pr)^2  {\bf \vec{Q}}_{lm \pm}^S( p r)  -  \left( (pr) {\bf \vec{b}}_{lm \pm}^S + {}_{(1)}{\bf \vec{b}}_{lm \pm}^S \right) e^{\pm i p r }\right] . 
\end{align}
In the first term, replacing the upper limit $\mu p$ by infinity leads to an error of order $1/p$, by the large argument asymptotics of ${\bf \vec{Q}}_{lm \pm}^S(v)$. A similar argument shows that the second term is of order~$1/p$. 
The third term is of order~$1/p$, by integration by parts~\cite{wong}, 
and a similar argument shows that the fourth term is of order $1/p^2$. 
In the fifth term, the expression in the square brackets is bounded, and it tends to zero for each positive $r$ as $p\to\infty$, whereas the expression multiplying these square brackets under the integral is bounded. 
The integral multiplying the overall factor $1/p$ hence goes to zero as $p\to\infty$, by dominated convergence, and the whole term is of order $o(1/p)$. 
Collecting these observations, we have 
\begin{align}
{\bf \vec{H}}^S_{lm<}( p ) = 
\begin{cases}
{\displaystyle p \left( {\bf \vec{a}}_{lm+}^S +  {\bf \vec{a}}_{lm-}^S \right) 
\int_0^\mu dr \; r^2 \; f^S_{lm}(r) + {\cal O} (p^2)} 
& \text{for} \ p \rightarrow 0 , 
\\[1ex]
{\displaystyle A_{lm}^S  \int_0^{ \infty} dv  \;  {\bf \vec{Q}}_{lm}^S( v) + {\cal O}(1/p)} 
& \text{for} \ p \rightarrow \infty . 
\end{cases}
\label{H<asymptotic}
\end{align}

Consider then ${\bf \vec{H}}^S_{\pm lm>}( p ) $~\eqref{H>}. We may proceed similarly. 
The leading term in the asymptotic expansion at large $p$ is of order~$1/p$, 
as is seen using \eqref{fasymptotics} and \eqref{Qasymptotics} and integrating by parts~\cite{wong}. 
To obtain the asymptotic expansion at small $p$, we use 
\eqref{fasymptotics} and \eqref{Qasymptotics} to rewrite \eqref{H>} as 
\begin{align}
{\bf \vec{H}}^S_{\pm lm>}( p ) & = B^S_{lm}  \int^\infty_{ \mu p} dv  \;  {\bf \vec{Q}}_{lm \pm}^S( v) 
+ {}_{(1)}B^S_{lm} \, p 
\int^\infty_{ \mu p} dv  \;  \frac{{\bf \vec{Q}}_{lm \pm}^S( v)}{v}
\notag
\\
& \hspace{3ex}
+  p \int^\infty_\mu dr \; \left( r^2  f^S_{lm}(r) - B^S_{lm} - {}_{(1)}B^S_{lm}/r \right) \; {\bf \vec{Q}}_{lm \pm}^S( p r) . 
\end{align}
In the first term, replacing the lower limit $\mu p$ by zero leads to an error of order~$p$, by the small argument asymptotics of ${\bf \vec{Q}}_{lm \pm}^S(v)$. 
The second term is of order $p\log p$, again by the small argument asymptotics of ${\bf \vec{Q}}_{lm \pm}^S(v)$. 
In the third term, ${\bf \vec{Q}}_{lm \pm}^S( p r)$ is bounded and the factor $\left( r^2  f^S_{lm}(r) - B^S_{lm} - {}_{(1)}B^S_{lm}/r \right)$ has the large $r$ falloff $1/r^2$; the term is hence of order~$p$. 
Collecting these observations, we have 
\begin{align}
{\bf \vec{H}}^S_{lm>}( p ) 
= 
\begin{cases}
{\displaystyle B^S_{lm}  \int_0^{ \infty} dv  \;  {\bf \vec{Q}}_{lm}^S( v) + {\cal O}(p \log p)} 
& \text{for} \ p \rightarrow 0 , 
\\[2ex]
{\displaystyle {\cal O}(1/p)} 
& \text{for} \ p \rightarrow \infty . 
\end{cases}
\label{H>asymptotic}
\end{align}

Combining \eqref{H<asymptotic} and~\eqref{H>asymptotic}, we find 
\begin{align}
{\bf \vec{H}}^S_{lm}( p ) 
= 
\begin{cases}
{\displaystyle B^S_{lm}  \int_0^{ \infty} dv  \;  {\bf \vec{Q}}_{lm}^S( v) + {\cal O}(p \log p)}  
& \text{for} \ p \rightarrow 0 , 
\\[2ex]
{\displaystyle A^S_{lm}  \int_0^{ \infty} dv  \;  {\bf \vec{Q}}_{lm}^S( v) + {\cal O}(1/p)} 
& \text{for} \ p \rightarrow \infty , 
\end{cases}
\label{Hasymptotic}
\end{align}
and $\alpha_r(0,{\bf p})$ \eqref{alphageneral} takes the form 
\begin{align}
\alpha_r(0,{\bf p}) = \frac{i }{p \sqrt{2 \omega_p}} & \sum_{l,m,r, S} \, {\bf \vec{H}}^S_{lm}( p )  \cdot \pol . 
\label{alphageneral-rev}
\end{align}

\subsection{The response function: the result}

For the response function~\eqref{F-infinite}, we need the pull-back 
of the two-point function $W_{A}(x^{\prime \prime}, x^{\prime})$ \eqref{Wasimp} to the inertial trajectory~\eqref{inertialtraj}, and this involves an integration over~${\bf p}$. 
While the formulas for $\alpha_r(0,{\bf p})$ in \eqref{Hasymptotic} and \eqref{alphageneral-rev} were obtained for 
${\bf p}$ oriented in the ${\bf z}$ direction, these formulas generalise to general ${\bf p}$ by allowing the coefficients $B^S_{lm}$ and $A^S_{lm}$ in \eqref{Hasymptotic} to depend on the direction of~${\bf p}$, and we now consider this done. 

For $W_{A}(x^{\prime \prime}, x^{\prime})$ \eqref{Wasimp}, 
the angular integral in the exponent has the form 
\begin{align}
\int d\Omega_p \, | \alpha_r(0,{\bf p})|^2 = \frac{1}{2 \omega_p^3} \int d\Omega_p \left|  \sum_{l,m,S} \, {\bf \vec{H}}^S_{lm}( p )  \cdot \pol \right|^2 , 
\label{alphasquared}
\end{align}
using~\eqref{alphageneral-rev}. Summing over $r$ and using~\eqref{Hasymptotic}, 
we hence have 
\begin{align}
W_{A}(\tau^{\prime \prime}, \tau^{\prime})  = \exp \! \left[  -\int_0^\infty dp \; p^2 \, h(p)\left( 1 - e^{-i p \left( \tau^{\prime \prime} - \tau^{\prime} \right) } \right) \right] , 
\label{Wagen}
\end{align}
where 
\begin{align}
h(p) = \sum_{r} \int \frac {d\Omega_p}{(2 \pi)^3} \, | \alpha_r(0,{\bf p})|^2 = 
\begin{cases}
{\displaystyle p^{-3} \bigl( \langle \chi_{1}  \rangle + {\cal O}(p \log p) \bigr)}
& \text{for} \ p \rightarrow 0 , 
\\[1ex]
{\displaystyle p^{-3} \bigl( \langle \chi_{2}  \rangle + {\cal O}(1/p) \bigr)}
& \text{for} \ p \rightarrow \infty , 
\end{cases}
\label{alphaasymptotic}
\end{align}
\begin{subequations}
\label{chi}
with 
\begin{align}
\langle \chi_{1}  \rangle & = \int \frac {d\Omega_p}{ 2 (2 \pi)^3} \sum_r \left|  \sum_{l,m,S} \, B^S_{lm}  \int_0^{ \infty} dv  \;  {\bf \vec{Q}}_{lm}^S( v)   \cdot \pol \right|^2 , 
\\
\langle \chi_{2}  \rangle & = \int \frac {d\Omega_p}{2 (2 \pi)^3} \sum_r \left|  \sum_{l,m,S} \, A^S_{lm}  \int_0^{ \infty} dv  \;  {\bf \vec{Q}}_{lm}^S(v)   \cdot \pol  \right|^2 . 
\end{align}
\end{subequations} 
Note that the small $p$ behaviour of $h(p)$ is determined by the large $r$ behaviour of the photon field through the coefficients~$B^S_{lm}$, and conversely the large $p$ behaviour of $h(p)$ is determined by the small $r$ behaviour of the photon field through the coefficients~$A^S_{lm}$; this is as one would have expected by the properties of Fourier transforms. Note also that the $S=Y$ term in the sums comes from the 
asymptotic \emph{radial\/} electric field, and is hence associated with soft charges analogous to those \eqref{softcharge1} and \eqref{softcharge2} in Section~\ref{special}.

As in Section~\ref{special}, we regulate the ultraviolet divergent integral over $p$ in \eqref{Wagen} 
by introducing an ultraviolet cut-off $\Lambda>0$. 
We show in Appendix \ref{B} that the regulated two-point function is given by 
\begin{align}
W_{A}(\tau^{\prime \prime}, \tau^{\prime}) 
= 
\exp \! \left\{  - \langle \chi_1 \rangle 
\log{\left[ \epsilon + i (\tau^{\prime \prime} - \tau^{\prime}) \right]} 
+ G\bigl(\epsilon + i (\tau^{\prime \prime} - \tau^{\prime})\bigr) - K_\Lambda\right\} , 
\label{Wainertial2gen}
\end{align}
where 
\begin{subequations}
\begin{align}
G(z) & = 
\int_0^\infty dp \;  \left(\frac{ p^3 h(p) - \langle \chi_{1}  \rangle}{p}  \right)  e^{-zp} , 
\\
K_\Lambda & = 
\langle \chi_{1}  \rangle
\left( 
\gamma + \log{\Lambda} 
\right) 
+ \int_0^\Lambda dp \;  \left(\frac{ p^3 h(p) - \langle \chi_{1}  \rangle}{p}  \right) , 
\end{align}
\end{subequations} 
and the limit $\epsilon \to 0^+$ is understood. 
$G(z)$ is a function of a complex variable, defined for $\Realpart z \ge 0$ except at $z=0$, holomorphic for $\Realpart z > 0$, and with the large $|z|$ falloff 
$G(z) = {\cal O}\!\left(\frac{\log z}{z}\right)$. 
The constant $K_\Lambda$ diverges at $\Lambda\to\infty$ as $\langle \chi_{2} \rangle \log{\Lambda}$: as the divergent part depends only on $\langle \chi_{2}  \rangle$, 
the divergence is fully determined by the short distance behaviour of the dressing.
Note that the result \eqref{Wainertial2} is recovered as the special case for the dressing considered in Section~\ref{special}. 

Using \eqref{F-infinite} with \eqref{Wphiinertial} and~\eqref{Wainertial2gen}, the response function on the inertial trajectory becomes 
\begin{align}
F(E)
&= \int_{-\infty}^{\infty} ds \; e^{-i E s} \; W_\phi (s) \; W_A (s) 
\nonumber \\
&= \frac{1}{4 \pi^2} \lim_{\epsilon \rightarrow 0^+} \int_{-\infty}^{\infty} ds \;  \frac{e^{-i E s + G(\epsilon + is)} \, e^{- K_\Lambda}}{\left( \epsilon + is  \right) ^{2+ \langle \chi_1 \rangle}} \; . 
\label{transrateinertial-gen}
\end{align} 
For $E>0$, a contour integral argument shows that $F(E) = 0$: 
as in Section~\ref{special}, the soft charges do not produce spontaneous excitations in the detector. 
The de-excitation rate, by contrast, does depend on the soft charges, 
in a way encoded in the function $G$ and the cutoff-dependent constant~$K_\Lambda$.

\section{Discussion\label{sec:discussion}}

We have addressed whether a soft photon cloud surrounding a charged particle can affect suitably localised quantum measurement processes involving the charged particle. Using Dirac's construction of gauge invariant field operators, we have considered a spatially pointlike two-level quantum detector, linearly coupled to the gauge invariant field operator of a charged scalar field in flat spacetime. 
The gauge invariant construction indirectly couples the quantum detector also to the electromagnetic field, 
through the photon cloud dressing that surrounds the charged particle excitation in the field. 

For a quantum detector on an inertial trajectory, we found that 
the excitation rate vanishes, independently of the dressing, but the 
the de-excitation rate depends on the soft charges
that correspond to the radial component of the electric field dressing at the asymptotic infinity. 
As the soft charges are a measure of the soft photons in the electric field, 
we conclude that the de-excitation rate of the local quantum detector is indeed sensitive to soft photon dressings.

The novelty in our quantum detector construction is the gauge invariant charged scalar field operator, instead of its non-gauge invariant counterpart, which is customarily used in quantum detector systems. 
In brief, our construction offers a natural way to define the total Lagrangian in a gauge invariant manner. 
One could argue that irrespective of the gauge invariant or non-invariant form of the field operator used in the interaction Hamiltonian, the physical quantities to be measured should not depend of the gauge chosen. However, the similarity between the Aharonov-Bohm effect and the electromagnetic memory \cite{Mao, Susskind, Bachlechner} suggests that explicit gauge dependence could manifest itself in other physical phenomena connected to the memory effect, soft charges and soft particles. The gauge invariant construction in the present paper then ensures that any spurious gauge effects other than the Aharonov-Bohm type are naturally eliminated. However, introducing the gauge invariant operator in the Lagrangian then makes it imperative to take into account the electric field dressing along with the charged particles, the consequence of which eventually result in the final expressions to be dependent on the corresponding soft charges as well.

\section*{Acknowledgements}

JL~acknowledges partial support by 
United Kingdom Research and Innovation (UKRI) 
Science and Technology Facilities Council (STFC) grant ST/S002227/1 
``Quantum Sensors for Fundamental Physics'' 
and Theory Consolidated Grant ST/P000703/1.

\appendix

\section{Appendix: Ultraviolet regularisation for $1/r^2$ dressing\label{A}}

In this appendix we verify the expression \eqref{Wainertial2} for the ultraviolet-regularised photon field contribution to the two-point function for the $1/r^2$ dressing of Section~\ref{special}. 

In the unregularised expression~\eqref{Wainertial}, the momentum integral in the exponent is ultraviolet divergent. Introducing an ultraviolet cutoff $\Lambda>0$, the exponent becomes a multiple of
\begin{align}
H_\Lambda(s) &= \int_0^\Lambda \frac{1 -  e^{-i p s }}{p} \, dp \nonumber \\
&= \int_0^\Lambda \frac{1 -  \cos(ps)}{p} \, dp
+ i \int_0^\Lambda \frac{\sin(ps)}{p} \, dp , 
\label{eq:HLambda-def}
\end{align}
where we have written $s = \tau''-\tau'$. 
Assuming $s\ne0$, the integrals can be written in terms of the sine and cosine integrals~\cite{DLMF}, with the result 
\begin{align}
H_\Lambda(s) &= \Cin(|s|\Lambda) 
+ i \sgn(s) \Si (|s|\Lambda) 
\notag
\\
&= - \Ci ( |s|\Lambda) + \log( |s|\Lambda) + \gamma
+ i \sgn(s) \left( \tfrac12 \pi + \si (|s|\Lambda) \right) , 
\end{align}
where $\gamma$ is the Euler-Mascheroni constant. 
Since both $\Ci(z)$ and $\si(z)$ fall off proportionally to 
$1/z$ at large positive argument, we take the regularised version of $H$ to be 
\begin{align}
H^{reg}_\Lambda(s) &= \log( |s|\Lambda) + \gamma
+ \frac{i\pi}{2} \sgn(s) 
\notag
\\
&= 
\gamma + \log{\Lambda} + \lim_{\epsilon \rightarrow 0^+} \log{\left( \epsilon + i s \right)} , 
\end{align}
where in the last expression $\log$ denotes the branch that is real-valued at positive argument. 
This gives~\eqref{Wainertial2}.

\section{Appendix: Ultraviolet regularisation for power-law dressing\label{B}}

In this appendix we verify the expression \eqref{Wainertial2gen} for the ultraviolet-regularised photon field contribution to the two-point function for the asymptotic power-law dressing of Section~\ref{general}. 

Introducing an ultraviolet cutoff $\Lambda>0$, the exponent in \eqref{Wagen} becomes the negative of
\begin{align}
Z_\Lambda(s)  = \int_0^\Lambda dp \; p^2  \; h(p) \left( 1 - e^{-i p s} \right) , 
\label{eq:ZLambda-def}
\end{align}
where we recall that $h(p)$ has the large and small $p$ asymptotics given by~\eqref{alphaasymptotic}. 
Note that if we view $s$ as a complex-valued variable, $Z_\Lambda(s)$ is well defined for $\Imagpart(s) \le 0$, 
and it is holomorphic for 
$\Imagpart(s) < 0$. 

We decompose $Z_\Lambda(s)$ as 
\begin{align}
Z_\Lambda(s)  &= \langle \chi_{1}  \rangle \int_0^\Lambda dp \;  \left( \frac{ 1 - e^{-i p s} }{p} \right)
+ \int_0^\Lambda dp \;  \left(\frac{ p^3 h(p) - \langle \chi_{1}  \rangle}{p}  \right) 
\notag \\[2ex]
&  \hspace{3ex} 
+  \int_\Lambda^\infty dp \;  \left(\frac{ p^3 h(p) - \langle \chi_{1}  \rangle}{p}  \right)  e^{-i p s} 
-  \int_0^\infty dp \;  \left(\frac{ p^3 h(p) - \langle \chi_{1}  \rangle}{p}  \right)  e^{-i p s} . 
\label{eq:ZLambda-2nd}
\end{align}
The first term in \eqref{eq:ZLambda-2nd} is a multiple of \eqref{eq:HLambda-def} and can be treated similarly. 
The second term is independent of~$s$. 
The third term falls off at large $\Lambda$ proportionally to $1/\Lambda$, 
uniformly in~$s$, by the large $p$ asymptotics of~$h(p)$. 
Finally, the fourth term is independent of $\Lambda$, 
and it has the large $|s|$ falloff ${\cal O}\!\left(\frac{\log (is)}{is}\right)$: 
the contribution to this falloff from large $p$ is ${\cal O}\!\left(\frac{1}{s}\right)$, 
by the large $p$ asymptotics of~$h(p)$, while the leading contribution comes from small~$p$, 
by the large $p$ asymptotics of~$h(p)$, on comparison with the identity 
\begin{align}
\int_0^\mu dp \log p \, e^{-zp} 
= 
- \frac{E_1(\Lambda z) + e^{-\mu z} \log\mu + \log z + \gamma}{z}  , 
\end{align}
valid for $z\ne0$ with $\Realpart z \ge 0$, where $\mu$ is a positive constant and $E_1$ is the exponential integral function~\cite{DLMF}. 

We hence take the regularised version of $Z_\Lambda$ to be 
\begin{align}
Z^{reg}_\Lambda(s) &= 
\langle \chi_{1}  \rangle \log{\left( \epsilon + i s \right)} 
- \int_0^\infty dp \;  \left(\frac{ p^3 h(p) - \langle \chi_{1}  \rangle}{p}  \right)  e^{-i p s} 
\notag\\
& \hspace{3ex}
+ \langle \chi_{1}  \rangle
\left( 
\gamma + \log{\Lambda} 
\right) 
+ \int_0^\Lambda dp \;  \left(\frac{ p^3 h(p) - \langle \chi_{1}  \rangle}{p}  \right) , 
\label{eq:Zreg-def}
\end{align}
where the limit $\epsilon \to 0^+$ is understood. 
The only $\Lambda$-dependence in $Z^{reg}_\Lambda(s)$ is in the additive constant, 
which diverges at $\Lambda\to\infty$ as $\langle \chi_{2}  \rangle \log{\Lambda}$. 
Note that as the divergent part depends only on $\langle \chi_{2}  \rangle$, 
it is fully determined by the short distance behaviour of the dressing. This is what one would have expected. 

This gives~\eqref{Wainertial2gen}.

\end{document}